\documentclass[a4paper,twoside]{article}
\usepackage[dvips]{graphicx}
\usepackage{latexsym,epsf}
\usepackage{amscd}
\usepackage{multicol}
\usepackage[centertags]{amsmath}
\usepackage{amsfonts}
\usepackage [utf8]{inputenc}
\usepackage [ukrainian, russian, english]{babel}
\usepackage {fancyhdr}

\newcommand{\bec}{\begin{center}}
\newcommand{\ec}{\end{center}}

\fancyfootoffset{1pt}%
\renewcommand {\footrule}{\vbox to 0pt{\hbox to \headwidth{ \hrulefill \hspace{63mm}}\vss}}

\makeatletter

\renewcommand{\ps@plain}{
\renewcommand{\@oddhead}{}
\renewcommand{\@evenhead}{}
\renewcommand{\@oddfoot}{\hfil \thepage}
\renewcommand{\@evenfoot}{\thepage \hfil \hfil}}
\makeatother 

\makeatletter
\renewcommand{\@biblabel}[1]{#1.\hfill}
\makeatother
\title{\textbf{\Large Investigations of electroweak symmetry breaking mechanism for Higgs boson decays into four fermions}}
\author{\textbf{\textit{T.V. Obikhod, I.A. Petrenko
\footnote{\normalfont
Corresponding author E-mail address: obikhod@kinr.kiev.ua} $\ \ $
}}
\\
\\
\emph{\small Institute for Nuclear Research, 
NAS of Ukraine 03028 Kiev, Ukraine}\\
{\small }}
\begin{document}
\selectlanguage{english}
\date{}
\maketitle

\thispagestyle{fancy}

\begin{center}
\begin{minipage}{165mm}
{\small
Models with extended Higgs boson sectors are of prime importance for
investigating the mechanism of electroweak symmetry breaking for Higgs decays
into four fermions and for Higgs-production in association with a vector bosons. In
the framework of the Two-Higgs-Doublet Model using two scenarios obtained from
the experimental measurements we presented next-to-leading-order results on the
four-fermion decays of light CP-even Higgs boson, $h \rightarrow 4f$. With the help of Monte
Carlo program Prophecy 4f 3.0, we calculated the values $\Gamma= \Gamma_{EW} /\left(\Gamma_{EW}+\Gamma_{SM}\right)$ and
$\Gamma= \Gamma_{EW+QCD} /\left(\Gamma_{EW+QCD}+\Gamma_{SM}\right)$ for Higgs boson decay channels $ H \rightarrow \nu_{\mu} \overline{\mu} e \overline{\nu_e}$, $\mu \overline{\mu} e \overline{e}$, $e \overline{e} e \overline{e}$. We didn't find significant difference when accounting
QCD corrections to EW processes in the decay modes of Higgs boson.	
	Using computer programs Pythia 8.2 and FeynHiggs we calculated the
following values: $\sigma(VBH)BR(H\rightarrow ZZ)$ and $\sigma(VBF)BR(H \rightarrow WW)$ for VBF production processes, $\sigma(ggH)BR(H \rightarrow WW)$ and $\sigma(ggH)BR(H \rightarrow ZZ)$ for gluon fusion production process at 13
and 14 TeV and found good agreement with experimental data.

}

\end{minipage}
\end{center}
\begin{multicols}{2}
\begin{center}
\textbf{\textsc{1. Introduction}}
\end{center}
	The discovery in 2012 of the Higgs boson and the subsequent studies of its properties put its compatibility with the Standard Model (SM).
However, as the SM Higgs boson is a scalar particle, it has sensitivity to possible new physics scales, connected with physics beyond the Standard Model (BSM). So there are the arguments for expecting new physics, called supersymmetry (SUSY). 

	Models with extended Higgs boson sectors are of prime importance for investigating the mechanism of electroweak symmetry breaking (EWSB) for Higgs
decays into four fermions and for Higgs-production in association with a vector bosons (VBF). In the renormalizable SM theory are accurate predictions for the coupling of the Higgs boson to all known particles, which influence the rates and kinematic properties of decay of the Higgs boson. Therefore, measurement of the decay rates and angular correlations yields information to probe the SM predictions for the Higgs boson.
 
		The observation of Higgs boson decays in four-lepton channels has provided an evidence that Higgs boson is responsible for VBF process through EWSB,  
$
\begin{CD}
p+p @> VBF >> H \rightarrow ZZ (WW)\rightarrow l\overline{l }l\overline{l}.
\end{CD}
$
 For the presented process we must take into account the following couplings:

    $\bullet$ define couplings for Higgs decaying to vector bosons;

    $\bullet$ define couplings for vector bosons decaying to fermions.
   
    Combined measurements of couplings $\kappa_W$ and $\kappa_Z$ in the decay of the Higgs boson are presented in Table 1, \cite{Aad_2020, Sirunyan_2019}. \\
\hspace*{-1mm}
\begin{minipage}{80 mm}	
\vspace{3 mm}	
\bec	
\emph{\textbf{Table 1.}} {\it Couplings modifier combined measurements from Run 1 and Run 2}	
\ec	
\bec	
\begin{tabular}{@{}lllll@{}}
\hline	
	
                                             & LHC Run 1 & ATLAS Run 2 & CMS Run 2 &  \\ 	
\hline	
$\kappa_\gamma$ & 0.87$\pm$ 0.14      & 1.05$\pm$ 0.09        & 1.07$\pm$ 0.09      &  \\	
$\kappa_W$      & 0.87$\pm$ 0.13      & 1.05$\pm$ 0.09        & -1.13$\pm$ 0.08     &  \\	
$\kappa_Z$      & -0.98$\pm$ 0.10     & 1.11$\pm$ 0.08        & 1.00$\pm$ 0.07      &  \\ \end{tabular}
\ec	
\vspace{0 mm}	
\end{minipage}

For the presented in Table 1 cases of negative values of the couplings they can be connected with quasi-
degenerate cases or with some new BSM effects.

	Therefore, measurements of couplings of Higgs boson to vector bosons are interesting to probe a model of BSM scenarios with new heavy particles contributing to the loops. It can be assumed that the new physics affecting the loops could introduce new decay channels i.e., may be branching ratio (BR) BSM are allowed. The experimental evidence of these processes is translated into constraints on the Higgs-boson width for decays into four-fermion final states, connected with measurements of BR and production cross sections.

	The total width of a 125 GeV SM Higgs boson is $\Gamma_H = 4.07 \times 10^{-3}$ GeV. But direct constraints on the Higgs boson width are much larger than the expected width of the SM Higgs boson. The intrinsic mass resolution in channel is about 1-2 GeV, so only upper limits on the Higgs boson width have been set by and the results are reported in Table 2, \cite{Aad_2014, CMS-PAS-HIG-16-041}.
\hspace*{-3mm}
\begin{minipage}{80 mm}
\vspace{3 mm}
\bec	
\emph{\textbf{Table 2.}} {\it  Run 1 observed (expected) direct 95\% CL constraints on the width of the
125 GeV resonance from fits to the $\gamma\gamma$ and ZZ mass spectra. The CMS measurement
from the 4l mass line-shape was performed using Run 2 data.}
\ec	
\bec	
\begin{tabular}{@{}lll@{}}\hline

Experiment         & $M_{\gamma\gamma}$  & $M_{4l}$   \\ \hline
ATLAS 	 & $< 5.0(6.2) GeV$       & $< 2.6(6.2) GeV$         \\
CMS      & $< 2.4(3.1) GeV$       & $< 1.1 (1.6) GeV$       \\
\end{tabular}
\ec
\vspace{0 mm}
\end{minipage}
	As BR is the function of Higgs boson mass \cite{group2013handbook,florian2016handbook} and decay width is not measured precisely, the
theoretical modeling and computer simulations of these observables are of importance for further investigations of Higgs boson properties and searches for BSM physics.

The article is devoted to the study of the properties of the Higgs boson through calculations of decay into four fermions and production cross sections, BR for the most probable channels of Higgs boson formation (gluon-gluon and VBF) and decay (ZZ, WW).

\begin{center}
\textbf{\textsc{2. Consideration of Higgs boson decay within THDM model}}
\end{center}
In the framework of the Two-Higgs-Doublet Model (THDM) \cite{Denner_2018} with the help of Monte Carlo program Prophecy 4f 3.0 \cite{denner2019prophecy4f} we have calculated NLO decay width for the following processes:
$$
H(p)\to f_1(k_1) + \overline{f_2}(k_2) + f_3(k_3) + \overline{f_4}(k_4)
$$
where $f$ and $\overline{f}$ denote fermion and anti-fermion with corresponding momenta 
$k_i$. To determine whether Higgs-like particle is SM particles is necessary to study the high-energy scattering amplitudes.

	The Higgs potential for two Higgs doublets $\Phi_1$ and $\Phi_2$ (hypercharge Y=-1 and 1) to generate down-type quarks/charged leptons ( $\Phi_1$ ) and up-type quarks ( $\Phi_2$) is given by
\begin{multline*}
V  = m^2_1 \Phi_1^{\dag}\Phi_1 + m^2_2\Phi_2^{\dag}\Phi_2 - m^2_3 \left( \Phi^T_1 i \sigma_2 \Phi_2 + h.c. \right) \\
 + \frac{1}{2}\lambda_1 \left( \Phi_1^\dag\Phi_1 \right)^2
 + \frac{1}{2}\lambda_2 \left( \Phi_2^\dag\Phi_2 \right)^2 \\
 + \lambda_3 \left( \Phi_1^\dag\Phi_1 \right) \left( \Phi_2^\dag\Phi_2 \right)
 + \lambda_4 \left| \Phi_1^Ti\sigma_2\Phi_2 \right|^2 \\
 + \frac{1}{2}\lambda_5 \left[ (\Phi_1^Ti\sigma_2\Phi_2)^2 + h.c. \right] \\
 + \left[ \left[ \lambda_6 (\Phi_1^\dag\Phi_1) + \lambda_7 (\Phi_2^\dag\Phi_2) \right]\Phi^T_1 i\sigma_2\Phi_2 + h.c. \right]
\end{multline*}

where $m^2_i = \mu^2 + m^2_{H_{i}}$ $(i = 1,2)$, with $\mu$ being the supersymmetric Higgsino mass parameter and $m_i$ the soft supersymmetri breaking mass parameters of the two Higgs doubletes; $m^2_3 \equiv B_\mu$ is assotiated to the B-term soft SUSY breaking parameter; 
$\lambda_i$, for $i$ = 1 to 7, are all the Higgs quartic couplings. After the spontaneous breaking of the EW symmetry, five physical Higgs particles are left in the spectrum: one charged Higgs pair, $H^{\pm}$, one CP-odd neutral scalar, $A$ and two $CP$-even neutral states, $H$ and $h$.
The vacuum expectation values of the neutral components $\Phi^0_i(i=1,2)$ of the two Higgs doubletes are equal to,

\begin{equation*}
\langle\Phi^0_i\rangle = \frac{\upsilon_i}{\sqrt{2}} 
\end{equation*}
 with $\tan \beta \equiv \frac{\upsilon_2}{\upsilon_1}$  and $\upsilon^2 = \upsilon_1^2 + \upsilon_2^2 = (246 \text{GeV})^2$.
	The Higgs sector depends on the electroweak gauge coupling constants and on the vacuum expectation value $\upsilon$ and is determined by only two free parameters: $\tan \beta$ - the ratio of the vacuum expectation values $\upsilon_2/\upsilon_1$ - and one Higgs boson mass, $CP$-odd Higgs boson mass, $m_A$. The phenomenology of the Higgs sector depends on the couplings of the Higgs bosons to gauge bosons and fermions. The couplings of the two $CP$-even Higgs bosons to W and Z bosons are given in terms of the angles $\alpha$, that diagonalises the $CP$-even Higgs boson squared-mass matrix, and $\beta$
\begin{equation}\label{h_eq}
\begin{gathered}
g_{hVV}=g_Vm_V\sin(\beta - \alpha), \\
g_{HVV}=g_Vm_V\cos(\beta - \alpha),
\end{gathered}
\end{equation}
where  $g_V=2m_V/\upsilon$, for $V=W \text{ or } Z$.\\
	Denote by $\kappa_i'(i=f,V)$,  couplings of Higgs boson to the fermions  and gauge bosons. The Higgs couplings to gauge bosons  are given by (\ref{h_eq}).	

	Yukawa Lagrangian for THDM model  is given in the mass basis by formula
\begin{multline*}
\mathcal{L}_{\text{Yukawa}} = - \sum_{f = u,d,l} \frac{m_f}{\upsilon}
( 
\xi^f_h\overline{f}fh + \xi^f_H\overline{f}fH  \\
- i\xi^f_A\overline{f}\gamma_5 fA ) 
	- \lbrace \frac{\sqrt{2}V_{ud}}{\upsilon}\overline{u} ( m_u\xi^u_AP_L  \\
	+ m_d\xi^d_AP_R )d\text{H}^+
	 + \frac{\sqrt{2}m_1\xi^l_A}{\upsilon}\overline{\upsilon}_L1_RH^+ + h.c.  
	\rbrace
\end{multline*}
where $u,d,l$ stand generically for up-type quarks, down-type quarks, and charged leptons of all three generations, respectively, $P_{L,R}$ are the projection operators of th left- and right-handed fermions, and $V_{ud}$ denotes the appropriate element of the CKM matrix. We have Type-1 THDM:
\begin{multline*}
\begin{gathered}
\xi^u_h = \xi^d_h = \xi^l_h \equiv \xi_h = \frac{\cos\alpha}{\sin\beta} \\
\xi^u_H = \xi^d_H = \xi^l_H \equiv \xi_H = \frac{\sin\alpha}{\sin\beta}
\end{gathered}
\end{multline*} 
Therefore, couplings of Higgs boson to the fermions  and gauge bosons are the following:
$\kappa_f\equiv\xi_h$ and $\kappa_f'\equiv\xi_H$; $\kappa_V\equiv g_{hVV}$ 
and $\kappa_V'\equiv g_{HVV}$.

\begin{center}
\textbf{\textsc{3. Results of decay width calculations}}
\end{center}

The partial width of $H \to 4f$ calculated with Prophecy 4f 3.0 program can be split into WW, ZZ and their interference
\begin{multline*}
\hspace*{-3mm}\Gamma^{\text{Proph}}_{4f} =\Gamma_{H\rightarrow W^*W^*\to 4f} +\\
+ \Gamma_{H\rightarrow Z^*Z^*\to 4f} + \Gamma_{WW/ZZ-int},
\end{multline*}
where $\Gamma_{H\rightarrow W^*W^*\to 4f} = \\
=9\cdot \Gamma_{H\to \nu_e \overline{e} \mu\overline{\nu_{\mu}}} + 12\cdot \Gamma_{H\to \nu_e \overline{e}d\overline{u}} + 4\cdot \Gamma_{H\to u \overline{d}s\overline{c}},
$
\\
\\
$
\Gamma_{H\to Z^*Z^*\to4f} = 3\cdot \Gamma_{H\to \nu_e \overline{\nu_e}\nu_{\mu} \overline{\nu_{\mu}}} + 3\cdot \Gamma_{H\to e \overline{e}\mu \overline{\mu}} \\ 
+ 9\cdot \Gamma_{H\to \nu_e \overline{\nu_e}\mu \overline{\mu}} + 3\cdot \Gamma_{H\to \nu_e \overline{\nu_e}\nu_e \overline{\nu_e}} \\
+ 3\cdot \Gamma_{H\to e \overline{e}e \overline{e}} + 6\cdot \Gamma_{H\to \nu_e \overline{\nu_e}u \overline{u}} \\
+ 9\cdot \Gamma_{H\to \nu_e \overline{\nu_e}d \overline{d}} + 6\cdot \Gamma_{H\to u\overline{u}e \overline{e}}\\
+ 9\cdot \Gamma_{H\to d \overline{d}e\overline{e}} + 1\cdot \Gamma_{H\to u\overline{u}c\overline{c}}\\
+ 3\cdot \Gamma_{H\to d \overline{d}s \overline{s}}+ 6\cdot \Gamma_{H\to u \overline{u}s \overline{s}}\\
+ 2\cdot \Gamma_{H\to u \overline{u}u \overline{u}}+ 3\cdot \Gamma_{H\to d \overline{d}d \overline{d}},
$
\\
\\
$
\Gamma_{WW/ZZ-int} = 3\cdot \Gamma_{H\to \nu_e \overline{e}e \overline{\nu_e}} - 3\cdot \Gamma_{H\to \nu_e\overline{\nu_e}\mu \overline{\mu}} \\ 
- 3\cdot \Gamma_{H\to \nu_e\overline{e} \mu\overline{\nu_{\mu}}} + 2\cdot \Gamma_{H\to u \overline{d}d \overline{u}} \\
- 2\cdot \Gamma_{H\to u \overline{u}s \overline{s}} - 2\cdot \Gamma_{H\to u \overline{d}s \overline{c}}.
$

Using two scenarios obtained from the experimental measurements \cite{Denner_2018} we presented next-to-leading-order results on the four-fermion decays of light CP-even Higgs boson, $h\to 4f$. With the help of Monte Carlo program Prophecy 4f 3.0, we calculated the values $\Gamma= \Gamma_{EW} /(\Gamma_{EW}+\Gamma_{SM})$ and $\Gamma= \Gamma_{EW+QCD} /(\Gamma_{EW+QCD}+\Gamma_{SM})$ for Higgs boson decay channels: $H\to \nu_{\mu} \overline{\mu} e \overline{\nu_e}$, $H\to \mu \overline{\mu} e \overline{e}$, $H\to e \overline{e}e \overline{e}$. We didn't find significant difference when accounting QCD corrections to EW processes in the decay modes of Higgs boson. The results of our calculations are presented in Tables 3, 4, 5.
\begin{minipage}{80 mm}
\vspace{3 mm}
\bec
\emph{\textbf{Table 3.}} {\it  Calculation of decay widths within the THDM model for the decay channel $H\to \nu_\mu \overline{\mu} e \overline{\nu_e}$.
}
\ec
\bec
\begin{tabular}{@{}llll@{}}
\hline
                        & EW+QCD/EW  & $\Gamma$ \\
\hline
7-B1 	 & 0.00991      & 0.491    &  \\
7-B2     & 0.009296      & 0.475   &  \\  
5-B1 	 & 0.00981      & 0.4889    &  \\
5-B2 	 & 0.009099      & 0.4701    &  \\
SM 	 	 & 0.01025      & 0.5    &  \\
\end{tabular}
\ec
\vspace{0 mm}
\end{minipage}

	B1 and B2 scenarios for THDM model are taken from \cite{Denner_2018} with the corresponding renormalization schema parameters: 7 or 5 taken for calculations within Prophecy 4f 3.0 program. The values $Gamma$ are calculated according to the formulas, $\Gamma= \Gamma_{EW} /(\Gamma_{EW}+\Gamma_{SM})$ and $\Gamma= \Gamma_{EW+QCD} /(\Gamma_{EW+QCD}+\Gamma_{SM})$.
\begin{minipage}{80 mm}
\vspace{3 mm}
\bec
\emph{\textbf{Table 4.}} {\it  Calculation of decay widths within the THDM model for the decay channel $H\to \mu \overline{\mu} e \overline{e}$.
}
\ec
\bec
\begin{tabular}{@{}llll@{}}
\hline

                        & EW+QCD/EW  & $\Gamma$ \\ \hline

7-B1(2) 	 & 0.000232      & 0.490    &  \\
5-B1(2)      & 0.0002296     & 0.4874   &  \\  
SM 			 & 0.000241      & 0.5    &  \\
\end{tabular}
\ec
\vspace{0 mm}
\end{minipage}
\begin{minipage}{80 mm}
\vspace{3 mm}
\bec
\emph{\textbf{Table 5.}} {\it  Calculation of decay widths within the THDM model for the decay channels $H\to e \overline{e}e \overline{e}$ and $H\to \mu \overline{\mu} e \overline{e}$.
}
\ec
\bec
\begin{tabular}{@{}llll@{}}
\hline
                    				 & 7-B1			& SM  & $\Gamma$ \\
\hline
$H\to e \overline{e}e \overline{e}$ 				 & 0.000127      & 0.000133    &  0.49\\
$H\to \mu \overline{\mu} e \overline{e}$      & 0.000232      & 0.000241   &  0.49\\  
\end{tabular}
\ec
\vspace{3 mm}
\end{minipage}

\begin{center}
\textbf{\textsc{4. Calculations of the Higgs boson production cross sections}}
\end{center}
	The observation of the Higgs boson was a major step towards the understanding of the mechanism of EWSB. LHC Higgs signal strength measurements, electroweak precision measurements are connected with measuring of the cross sections times BR, $\sigma \cdot \text{BR}$. In the paper \cite{ATLAS-CONF-2018-018} was presented experimental data of production cross sections of the Higgs boson in proton-proton collisions in the $H \to Z Z \to 4l$ decay channel. The cross section is measured to be $\sigma = 4.04 \pm 0.47$ fb, while the SM prediction is $\sigma_{SM} = 3.35 \pm 0.15$ fb. The cross-section times $H \to Z Z$ BR for gluon fusion and VBF production are measured to be $1.22 \pm 0.18$ pb and $0.25 \pm 0.09$ pb, respectively. In spite of the fact that these measurements are in agreement with the SM predictions it was important for us to calculate these observables in the framework of THDM model for explanation of the deviations from experimental data. The calculations gave us the possibility to check the viability of BSM predictions with the corresponding range of parameters.
	
	With the help of FeynHiggs program \cite{feyn_higgs} and using the parameters $\tan\beta = 3$, $m_A=200$ we have calculated $BR(H\rightarrow WW) = 0.54407$, $BR(H\rightarrow ZZ) = 0.20721$. We used Pythia 8.2 program for calculation of VBF and top fusion production cross sections of $CP$-even Higgs boson. The corresponding production processes are presented in Fig.1, from \cite{wikiHiggs}.
\begin{minipage}{80 mm}
\vspace{1 mm}
\bec
\includegraphics[width=0.45\textwidth]{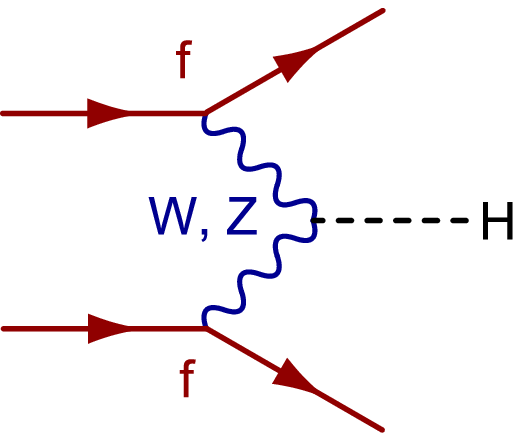}
\includegraphics[width=0.45\textwidth]{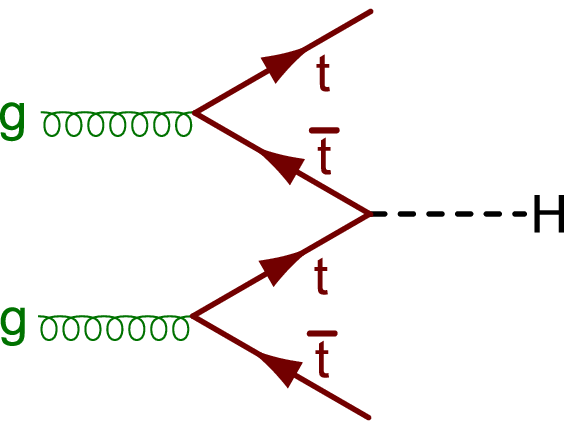}
\emph{\textbf{Fig.1.}} {\emph{ Feynman diagrams for Higgs production process through: left - VBF process, right - top fusion
process.}}\\
\ec
\vspace*{1mm}
\end{minipage}

	The results of our calculations are presented in Table 6 for 13 TeV energy of proton-proton collisions and in Table 7 for 14 TeV.
\begin{minipage}{80 mm}
\vspace{3 mm}
\bec
\emph{\textbf{Table 6.}} {\it  VBF and top fusion production cross sections of CP-even Higgs boson at 13 TeV.}
\ec
\bec
\begin{tabular}{@{}llll@{}}
\hline
E$_{cm}$=13TeV                & $\sigma$ (mb$^{-1}$) \\
\hline
$ff'\to H_0(H_2)ff' (ZZ)$ 	 		   & $3.047\cdot 10^{-10}$ &  \\
$f_1f_2\to H_0(H_2)f_3f_4 (W^+W^-)$    & $7.955\cdot 10^{-10}$ &  \\  
$gg \to H_0(H_2) t \overline{t}$ 	   & $1.906\cdot 10^{-11}$ &  \\
\end{tabular}
\ec
\vspace{0 mm}
\end{minipage}
	The cross-section times $H \to Z Z$ BR for gluon fusion and VBF production processes are measured to be $1.22 \pm 0.18$ pb and $0.25 \pm 0.09$ pb \cite{ATLAS-CONF-2018-018}, and calculated data are \\
	$\sigma(ggH)BR(H\to ZZ)= 1.04235$,\\
	$\sigma(VBF) \cdot BR(H\to ZZ)=0.16484.$\\
\begin{minipage}{80 mm}
\vspace{3 mm}
\bec
\emph{\textbf{Table 7.}} {\it VBF and top fusion production cross sections of CP-even Higgs boson at 14 TeV.}
\ec
\bec
\begin{tabular}{@{}llll@{}}
\hline
E$_{cm}$=14TeV                & $\sigma$ (mb$^{-1}$) \\
\hline
$ff'\to H_0(H_2)ff' (ZZ)$ 	 		   & $3.556\cdot 10^{-10}$ &  \\
$f_1f_2\to H_0(H_2)f_3f_4 (W^+W^-)$    & $9.337\cdot 10^{-10}$ &  \\  
$gg \to H_0(H_2) t \overline{t}$ 	   & $2.730\cdot 10^{-11}$ &  \\
\end{tabular}
\ec
\vspace{0 mm}
\end{minipage}
	The comparison with experimental data gives good agreement and the deviations are connected with uncertainty of parameter space.

	In the article \cite{Aaboud_2019}  experimental data were presented on the product of the $\sigma(ggH)\times$ BR$(H\to WW)$ for the gluon-gluon fusion production cross section to be $11.4\pm1.2\pm 1.8$pb and on the product of the 
	$\sigma(VBH)\times$ BR$(H\to WW)$ for VBF production cross-section to be $0.50\pm0.24\pm0.17$pb.
	
	Using code SusHi and the parameter space $\tan\beta=3$, $m_A=200$ GeV we calculated production cross sections for gluon-gluon fusion process at 13 TeV, 
	$\sigma(ggh) = 5.03$pb. As the calculated by FeynHiggs value of BR is 
	$BR(H \to WW)=0.54047$, we have the following results:\\
	$\sigma(ggH) \cdot BR(H\to WW)=2.72, $\\
	$\sigma(VBF) \cdot BR(H\to WW)=0.43. $\\
	We see good agreement of our calculations especially for the second case from comparison with experimental data.

\begin{center}
\textbf{\textsc{5. Conclusions}}
\end{center}

The study of the properties of the Higgs boson is still an open problem, connected both with experimental difficulties and  theoretical interpretation of its properties. Especially important is the study of the EWSB mechanism, which is associated with the formation of the Higgs boson through the VBF channel and the subsequent decay into four leptons. Since the experimental determination of the Higgs boson decay width for the presented decay channel is a difficult task and is represented only by the upper boundary of both the ATLAS and CMS collaborations, we decided to calculate the decay widths for the four leptonic Higgs boson decay channels in the framework of THDM model. The calculated results show a slight deviation from the SM calculations, but vary for different decay channels, especially for  channel $H\to \nu_\mu \overline{\mu} e \overline{e}$ with B2 scenario. In addition, the results of calculations of the values $\Gamma= \Gamma_{EW} /(\Gamma_{EW}+\Gamma_{SM})$ and $\Gamma= \Gamma_{EW+QCD} /(\Gamma_{EW+QCD}+\Gamma_{SM})$ taking into account the electroweak and QCD corrections indicate the absence of the effect of QCD corrections on the decay width for the presented decay channels of the Higgs boson.

	Another step towards the understanding of the mechanism of EWSB is the calculation of the Higgs boson production cross section $\times$ BR. We have considered ggH and VBF production modes of Higgs boson and the calculations were made with the help of Pythia8.2 and SusHi programs. Using FeynHiggs program with the same parameter region ($\tan\beta=3$, $m_A=200$ GeV) we
calculated $BR(H\to ZZ)$ and $BR(H\to WW)$ values. The obtained results are in good agreement with experimental data. We also calculated the same values for the
energy of 14 TeV for subsequent searches of Higgs signal at the LHC.

\bibliographystyle{ieeetr}

\end{multicols}

\end{document}